\documentclass[conference]{IEEEtran}
\IEEEoverridecommandlockouts
\usepackage{cite}
\usepackage{amsmath,amssymb,amsfonts}
\usepackage{algorithmic}
\usepackage{graphicx}
\usepackage{textcomp}
\usepackage[table,xcdraw]{xcolor}
\usepackage{tikz}
\usepackage{subcaption}
\usepackage[colorinlistoftodos,prependcaption]{todonotes}
\usepackage{pgfplots}
\usetikzlibrary{pgfplots.statistics}

\usepackage{xcolor}
\usepackage[ruled,vlined]{algorithm2e}
\usepackage{xcolor}
\usepackage{listings}
\usepackage{url}
\def\BibTeX{{\rm B\kern-.05em{\sc i\kern-.025em b}\kern-.08em
    T\kern-.1667em\lower.7ex\hbox{E}\kern-.125emX}}
    
\newcommand\copyrighttext{%
  \footnotesize \textcopyright 2021 IEEE. Personal use of this material is permitted.
  Permission from IEEE must be obtained for all other uses, in any current or future
  media, including reprinting/republishing this material for advertising or promotional
  purposes, creating new collective works, for resale or redistribution to servers or
  lists, or reuse of any copyrighted component of this work in other works.
  }
\newcommand\copyrightnotice{%
\begin{tikzpicture}[remember picture,overlay]
\node[anchor=south,yshift=10pt] at (current page.south) {\fbox{\parbox{\dimexpr\textwidth-\fboxsep-\fboxrule\relax}{\copyrighttext}}};
\end{tikzpicture}%
}

\begin{document}

\title{EdgePier: P2P-based Container Image Distribution in Edge Computing Environments

}

\author{\IEEEauthorblockN{Soeren Becker}
\IEEEauthorblockA{\textit{Distributed and Operating Systems} \\
\textit{TU Berlin}\\
Berlin, Germany \\
soeren.becker@tu-berlin.de}
\and
\IEEEauthorblockN{Florian Schmidt}
\IEEEauthorblockA{\textit{Distributed and Operating Systems} \\
\textit{TU Berlin}\\
Berlin, Germany\\
florian.schmidt@tu-berlin.de}
\and
\IEEEauthorblockN{Odej Kao}
\IEEEauthorblockA{\textit{Distributed and Operating Systems} \\
\textit{TU Berlin}\\
Berlin, Germany \\
odej.kao@tu-berlin.de}
}

\maketitle
\IEEEpubidadjcol
\copyrightnotice

\begin{abstract}
Edge and fog computing architectures utilize container technologies in order to offer a lightweight application deployment. Container images are stored in registry services and operated by orchestration platforms to download and start the respective applications on nodes of the infrastructure. 
During large application rollouts, the connection to the registry is prone to become a bottleneck, which results in longer provisioning times and deployment latencies.
Previous work has mainly addressed this problem by proposing scalable registries, leveraging the BitTorrent protocol or distributed storage to host container images. However, for lightweight and dynamic edge environments the overhead of several dedicated components is not feasible in regard to its interference of the actual workload and is subject to failures due to the introduced complexity. 

In this paper we introduce a fully decentralized container registry called EdgePier, that can be deployed across edge sites and is able to decrease container deployment times by utilizing peer-to-peer connections between participating nodes. Image layers are shared without the need for further centralized orchestration entities. The conducted evaluation shows that the provisioning times are improved by up to 65\% in comparison to a baseline registry, even with limited bandwidth to the cloud. 
\end{abstract}


%
\section{Introduction}
Container virtualization has massively gained in importance and prevalence over the last years due to its extensively usage in research and industry. In contrast to traditional virtualization,
containers offer a lightweight and flexible approach in which applications are packaged, delivered and deployed together with their needed dependencies. Through orchestration platforms such as Kubernetes, containers are utilized to automate, manage, and maintain container-based workloads. For instance, in 2018 Nextflix launched around half a million containers daily\footnote{https://netflixtechblog.com/titus-the-netflix-container-management-platform-is-now-open-source-f868c9fb5436} and Google claims to run every part of the technology stack in containers\footnote{https://speakerdeck.com/jbeda/containers-at-scale}.

The lightweight nature of container virtualization makes it a good fit for upcoming, highly distributed architectures such as Edge and Fog Computing environments \cite{ismailEvaluationDockerEdge2015b}. By leveraging Linux kernel namespaces and cgroups, the isolation of separate processes is enabled \cite{sharmaContainersVirtualMachines2016} while only creating a minimal overhead on the host operating system \cite{morabitoHypervisorsVsLightweight2015}.
In addition, the portability, reusability, and scalability of container technologies allows for a rapid application deployment in dynamic and possibly heterogeneous edge and fog environments, that can scale with the demand, quickly tear down and optimize cost efficiency \cite{bellavistaFeasibilityFogComputing2017b}.
 
Subsequently, containerization is an important building block for a variety of applications in the IoT domain, for instance in the area of Edge Intelligence where containers are used to deliver and deploy machine learning algorithms and models to remote edge sites in order to analyze and act on the ever increasing generated data in the edge-cloud continuum  \cite{al-rakhamiLightweightCostEffective2020}. 
Another use case is the serverless or Function-as-a-Service (FaaS) paradigm, which gains in importance as a workload execution model at the edge of the network \cite{hallExecutionModelServerless2019, gliksonDevicelessEdgeComputing2017}: Stateless functions are provided in containers and only started when requested, automatically scaled up and down based on the demand and stopped if they are no longer needed \cite{baldiniServerlessComputingCurrent2017a}.
Both application domains rely on a fast and efficient container image distribution, for instance to keep the applied AI models up-to-date to all nodes of an edge site or to cope with the cold-start problem of FaaS \cite{mannerColdStartInfluencing2018}.

Container images are typically provided in container registries and downloaded by the respective runtime agent in order to start a containerized application on a node of the infrastructure. The container registry is often a centralized service deployed in the cloud, i.e. the public Docker Hub or a private deployment. Since edge environments are distributed i.e. across a smart city and located far away from the cloud, the uplink connection is often affected by high latencies and low bandwidth.
Therefore, in case of centralized registries, the connection to the registry is prone to become a bottleneck \cite{kangjinFIDFasterImage2017}, resulting in longer deployment latencies in general, increased cold-start times for FaaS, and decreased model update times in AI use cases.

Current approaches in data center environments try to tackle this situation by providing scalable container registries to cope with the increasing traffic demand. They rely on either shared distributed storage between participating nodes \cite{zhengWharfSharingDocker2018, nathanCoMIConCoOperativeManagement2017,duCiderRapidDocker2017} or employ peer-to-peer technologies that are in most cases built on BitTorrent, to distribute the download traffic load between a set of nodes \cite{kangjinFIDFasterImage2017}. Although the approaches show promising results in cloud data centers in regards to deployment latencies, they require the deployment of further dedicated services and applications or intrusive changes in the container runtime to manage, orchestrate, and distribute images across devices. Consequently, they are challenging to apply in Edge/Fog computing environments, where resources are already limited and devices are often incorporated in dynamic ad-hoc architectures that are affected by node churns, unreliable Edge-Cloud uplink connections or network partitions.

Therefore, we propose EdgePier, a fully decentralized P2P container registry that is able to optimize the loading of container images from participating nodes in an edge site. Furthermore, images are distributed across all nodes of the site to achieve high-availability, and by distributing the download traffic between all nodes the Edge-Cloud uplink connection is retrieved and deployment latencies are decreased.
 

The key contributions of this paper are:
\begin{itemize}
    \item We outline container provisioning as a relevant deployment scenario in edge computing and define requirements for a registry service used in distributed edge sites.
    \item We design and propose a novel architecture and implementation of a decentralized peer-to-peer container registry called EdgePier.
    \item We conduct an empirical evaluation in which we compare our implementation with a baseline registry for different container deployment scenarios in edge computing. 
\end{itemize}

The remainder of this paper is structured as followed:
Section II provides an overview about Docker images and registries before Section III depicts use cases and requirements for the decentralized container registry. Afterwards we describe the architecture of EdgePier in Section IV and the conducted experiments as well as results (Sec. V). Finally, we present the related work (Sec. VI) and conclude the paper in Section VII.

%
\section{Background}
In this section we give an overview about the architecture of Docker images and container registries. Furthermore, we describe and highlight features of the InterPlanetary File System (IPFS), as it is one of the main building blocks of EdgePier.

\subsection{Docker Images}
Docker images consists of two main components, a \textit{manifest} and \textit{layer blobs}. An image layer is a read-only file system that was created from one instruction in the Dockerfile during the image building process \cite{kangjinFIDFasterImage2017}. Subsequently, it describes an intermediate state of the image and represents different software or dependencies installed during that stage \cite{duCiderRapidDocker2017}.

The compressed \textit{layers} are called blobs and listed in the \textit{manifest} which contains all layers an image consists of. The layered structure of Docker images also enables deduplication, which allows for the sharing of layers between images \cite{zhengWharfSharingDocker2018}: Layers which are used by several containers only need to be downloaded once on a node.

During a container start, the docker daemon first loads the manifest file and subsequently fetches all compressed layer blobs that are not yet available on the local node, before creating the container from the decompressed layers, adding a further, writeable container layer and configuring the network and Linux namespaces as well as cgroups \cite{littleyBoltScalableDocker2019a}.

\subsection{Image Distribution}
Docker images are stored in a container registry, which is a stateless RESTful API that responds to image pull and push requests from Docker daemons. After a Docker image was created, the compressed \textit{layer blobs} that are not yet available in the registry storage are uploaded together with the \textit{manifest} via a POST request \cite{littleyBoltScalableDocker2019a} by the local Docker daemon.
Docker relies on content-addressable images\footnote{https://docs.docker.com/registry/spec/manifest-v2-2/} by providing a SHA256 hash of the layers content. This hash identifier is listed in the manifest, used to request the specific layer and also utilized to verify downloaded blobs by the local docker daemon.

A container registry also enables the storage of different versions of a container image, indicated by an image tag. Depending on which image tag the Docker daemon requests, the registry provides the respective manifest to the local daemon which consequently retrieves all not yet available layers via GET requests by their SHA256 based hash identifier. 

\subsection{IPFS}
The InterPlanetary File System (IPFS) \cite{benetIPFSContentAddressed2014} is a peer-to-peer version controlled distributed filesystem.
IPFS uses the Kademlia Distributed Hash Table (DHT) \cite{maymounkovKademliaPeertoPeerInformation2002} to find the network addresses of peers as well
as IPFS objects hosted by those peers. Furthermore, IPFS offers the exchange of blocks between participating peers based on the BitSwap protocol which works
similiar to the BitTorrent protocol. This results in a high-throughput block storage solution which scales to a large number of nodes \cite{benetIPFSContentAddressed2014}.

In addition, IPFS adresses objects by a cryptographic representation of the content itself, called content identifier (CID). Each time an object is stored in
IPFS, a unique hash of the content is created which can be used by other peers to download this object. 
The benefits of this approach are that objects inside of IPFS are tamper resistant, peers can verify
the blocks they downloaded using the CID and objects inside of IPFS are location-independent - every peer hosting the object shares
it with other peers asking for it.

The data structure used by IPFS to store objects is a Merkle directed acyclic graph (DAG) \cite{10.1007/3-540-48184-2_32} which on the one hand offers the aforementioned tamper resistance
and verification but also provides deduplication of content: Objects with the exact same content will have the same hash and therefore only need to
be stored once in the DAG.

Summarizing, IPFS offers a modular network layer based on the libp2p protocol \cite{benetIPFSContentAddressed2014} which is able to run on top of any network technology, uses a DHT to store
object locations and metadata and deploys an Object Merkle DAG on top to enabled content adressing, tamper resistance and deduplication of data. Peers are
able to exchange blocks with each other without the need of a centralized entity.

%
%
%
\section{Requirements} 
\begin{figure*}[ht]
    \centering
    \includegraphics[width=0.9\textwidth]{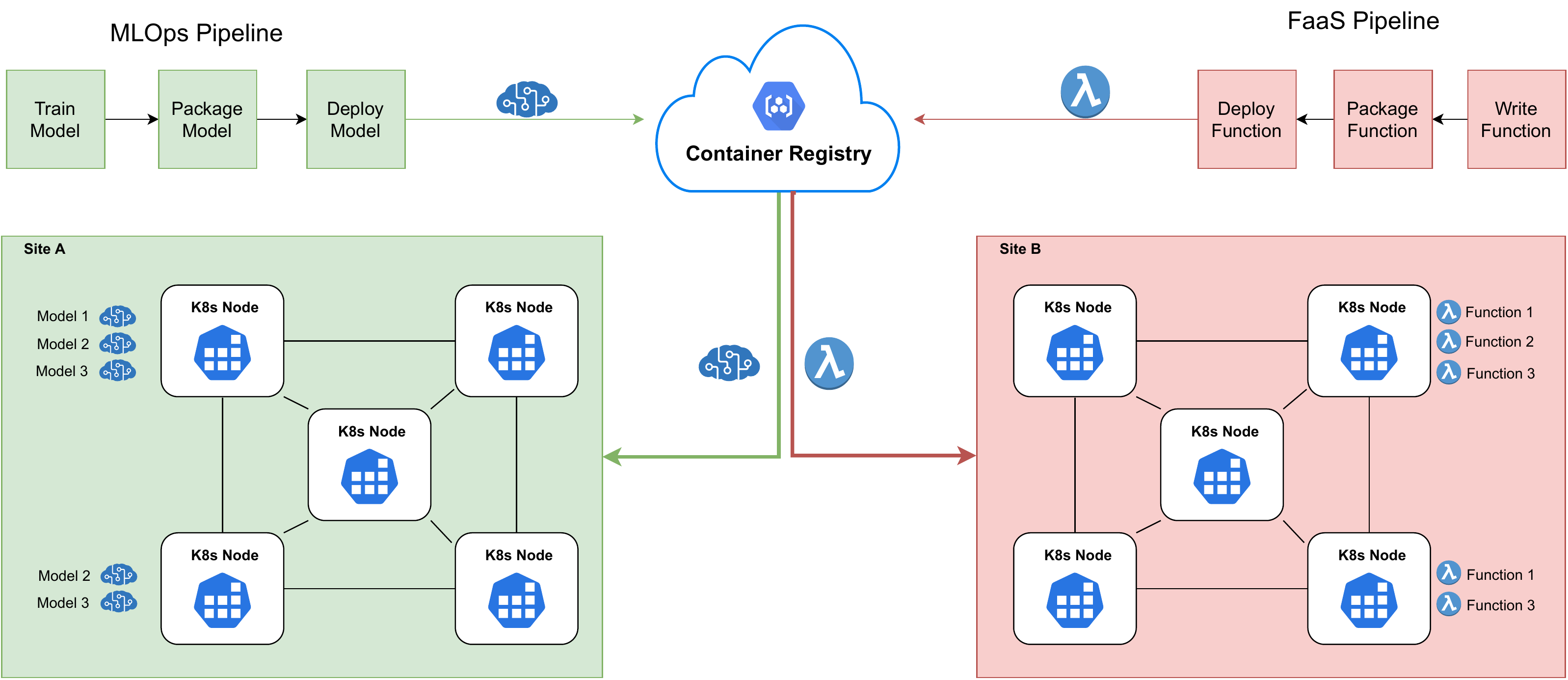}
    \caption{An edge computing environment consisting of two edge sites and a container registry located in the cloud. Site A is used for a machine learning use case, whereas the nodes in Site B are utilized by a Function-as-a-Service framework. MLOps and FaaS pipelines located in the cloud create container images which are subsequently deployed to the respective sites.}
    \label{fig:environment}
\end{figure*}

Deployments in the Edge Cloud Continuum are typically constructed like depicted in Figure \ref{fig:environment}: Several edge sites are located in swarms across for instance a smart city.
The edge devices inside of a site are often interconnected with high bandwidth and low latencies but the uplink connection to the cloud can be affected by high latencies, low bandwidth, packet loss, or other network anomalies, due to the remote location. 
In addition, Figure \ref{fig:environment} presents two common use cases for Edge Computing: On the one hand, edge sites are often used for AI scenarios, where ML models are applied on edge devices to i.e. monitor traffic behaviour, temperature, or other sensors \cite{al-rakhamiLightweightCostEffective2020}. The model training is commonly offloaded to the cloud, since lightweight edge devices are not always able to execute resource intensive training tasks without interfering with other workload on the device.

For that reason, MLOPs pipelines that train, optimize, validate, and package the model are deployed on cloud data center servers and often supported by hardware accelerators. The trained model is finally stored in a model repository, for instance as a container in a container registry, from where it gets deployed to different edge sites.

The second use cases, depicted in Figure \ref{fig:environment}, is the aforementioned Function-as-a-Service scenario, which enables developers to provide function snippets in the programming language of their choice. The functions are packaged -- together with needed dependencies and libraries -- in a container and uploaded to a container registry.
FaaS frameworks deployed in edge sites subsequently pull the containerized function when requested and create as well as execute function replica containers based on the demand \cite{gliksonDevicelessEdgeComputing2017}.

Similar to the aforementioned particular use cases, edge application in general are also often tested, compiled, and packaged in continuous integration pipelines deployed in cloud environments. 

Therefore, for the remainder of this paper we assume an environment in which applications are packaged in container images and uploaded to a container registry service located in the cloud. In addition, several edge sites deploy the applications by downloading the images from the registry service and starting the respective containers. 
As a further assumption, all devices inside a swarm are generally employed for a common purpose, i.e. as illustrated in Figure \ref{fig:environment} to apply similar AI models or execute FaaS workload.

In order to facilitate and improve the image deployment for such highly distributed environments, we establish the following requirements for a container registry service: 

\paragraph{Non-intrusive deployment}
The registry should work without the need to adapt the container runtime, to be able to applicable in a wide range of edge deployments and applications.


\paragraph{Reliable and fully decentralized}
Images should be synchronized across registry services inside an edge site, without the need for centralized orchestration or metadata services. In addition, the registry needs to be able to seamlessly cope with unreliable network connections, node join, churns  or even network partitions \cite{hoqueContainerOrchestrationFog2017a}.

\paragraph{Accelerated image distribution}
The image distribution process for remote edge sites should be accelerated compared to a central registry -- especially under unreliable network conditions -- to warrant the deployment of a P2P image registry.

\section{EdgePier Architecture}
This section describes the design and implementation of the EdgePier prototype. We give an overview about the used technologies, the design and implementation of the peer-to-peer registry based on the previously stated requirements.

\subsection{EdgePier Overview}
The EdgePier registry is a service which is envisioned to run on each node of an edge site to enable local pulling of docker images. The architecture consists of two main components: In order to answer pull requests by the docker daemon, EdgePier utilizes an HTTP API which is compatible with the official Docker registry API and automatically translates and forwards image manifest and layer requests to the IPFS network \footnote{https://github.com/ipdr/ipdr}. Due to the compatible API, the docker or kubelet daemon code does not 
need to be adjusted to allow downloading from the IPFS network which subsequently allows for an non-intrusive integration of the registry into pre-existing environments.

As a second component, we provide an IPFS agent alongside the registry API that is used to download requested images from IPFS peers. The agent is coupled with a configurable amount of local storage in which downloaded images are cached and consequently provided to other peers requesting the same image. 

To enable a high-available storage of docker images, we further integrated a replication strategy into the EdgePier prototype: In case one of the EdgePier registries is utilized by a container runtime agent to download an image which is not yet available in the respective edge site, it is automatically cached in the local IPFS storage of other EdgePier registries in the site. This is accomplished by employing the ipfs-cluster service\footnote{https://cluster.ipfs.io/} that maintains a global set of replicated files across a set of IPFS nodes. One of the assumptions for EdgePier is, that nodes inside an edge site are used for a common purpose and therefore often deploy the same container images. Thus, in the default configuration of EdgePier images are replicated to each node of the site, although the replication factor is configurable and images could also be stored on only i.e. three random nodes of the site. 



\subsection{Image Pulling Workflow}
\begin{figure}[h]
    \centering
    \includegraphics[width=0.45\textwidth]{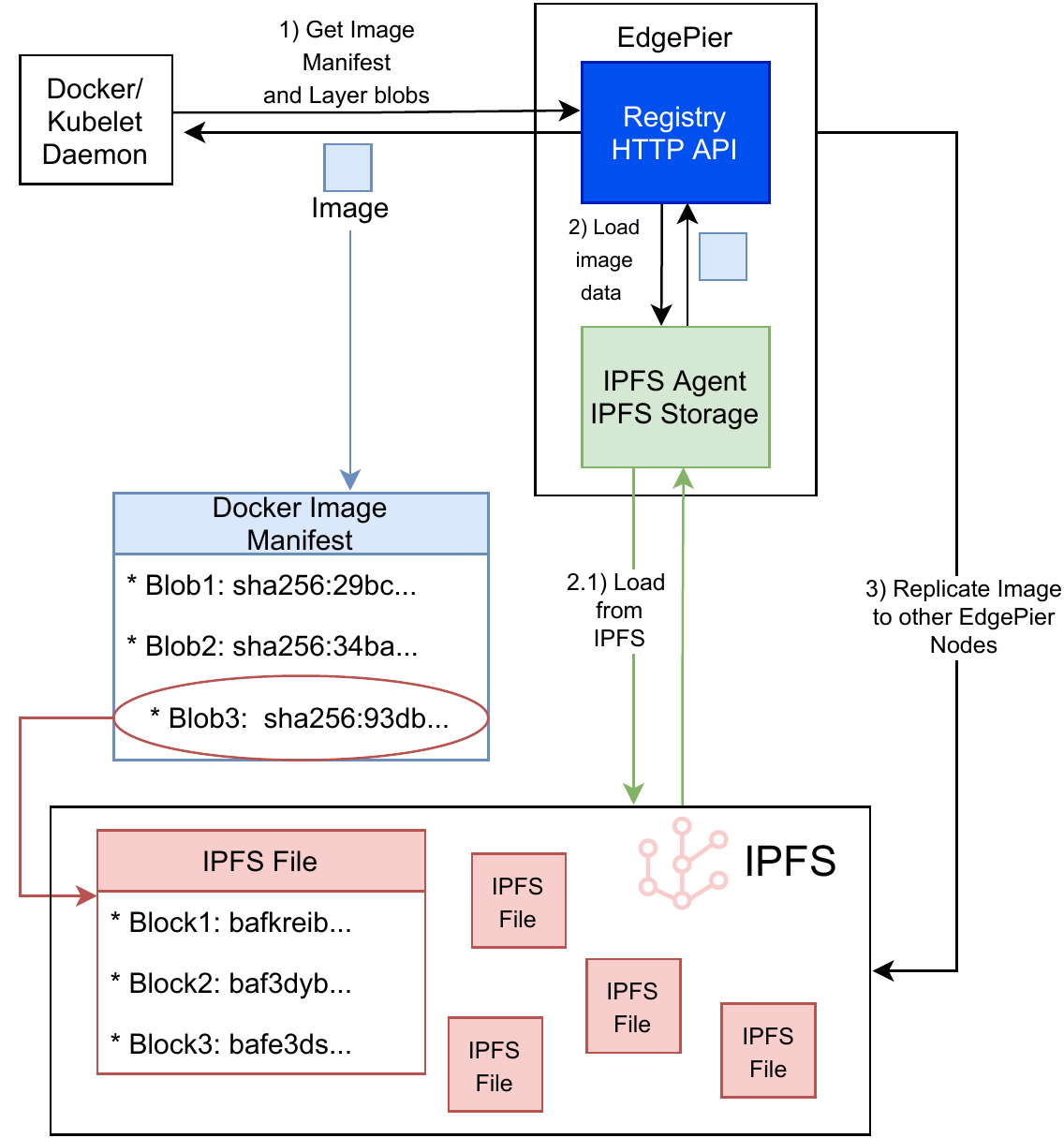}
    \caption{Image pulling workflow with the EdgePier registry.}
    \label{fig:edgepier}
\end{figure}
The image pulling process works like depicted in Figure \ref{fig:edgepier}:
In case a Pod is scheduled on a Kubernetes node of an edge site, the local kubelet daemon attempts to pull the respective image from the EdgePier registry that is deployed on the same node. At first, EdgePier tries to load the Image Manifest from the local IPFS storage, which contains links to all layers of the actual image. After the the kubelet daemon receives the image manifest, it requests all listed layer blobs, again from the local EdgePier registry. If the image manifests and layer blobs are already available (cached) in the local IPFS storage, they 
are directly provided to the kubelet daemon via the HTTP API. In case of a cold start, meaning the respective 
image data is not available in the local storage because the image was not yet deployed in the site, EdgePier locates the respective image data in the IPFS network and downloads it from each peer currently providing the files. 

As shown in Figure \ref{fig:edgepier}, each layer of a docker image refers to a file in the IPFS network which in turn is splitted into blocks that are distributed across peers in the IPFS network.
Similar to the BitTorrent protocol, the download traffic is distributed over all IPFS peers that store some of the blocks in their local storage. Consequently, the image download speed increases with the amount of peers caching the image data. At the same time, the Edge-Cloud uplink to a traditional image registry is relieved since other edge devices or sites -- located in closer proximity -- are leveraged for the download traffic. Furthermore, the single-point-failure is eliminated since images are available as long as at least one IPFS peer still stores the respective data and can be downloaded even in case of a network outage on the uplink layer.

\subsection{Deployment}
In order to fulfill the \textit{non-intrusive deployment} requirement, EdgePier was designed to enable a straightforward usage in orchestration platforms: The registry can be installed as a single docker container and integrates without any adjustments to the underlying container runtime. EdgePier needs to be configured as a private registry in i.e. Kubernetes so that kubelet daemon refers to the local instance on the node.

Since the registry relies on IPFS as a storage backend, image data can be located, downloaded and shared using the decentralized libp2p protocol. Consequently, no central metadata or orchestration service is needed for the provisioning of images. The container images can be pulled as long as at least one IPFS peer still has the data available and in case a new node joins the site, the currently used images in that edge site are automatically synchronized using ipfs-cluster. Therefore, the \textit{reliable and fully decentralized} requirement is also fulfilled.



%
%
\section{Evaluation}
We conducted several experiments to evaluate the feasibility and performance of EdgePier for container image provisioning to remote edge sites. In this section, we first give an overview about the used test bed and configuration before describing and evaluating two different container distribution scenarios. Finally, we present the impact of the image size on the provisioning time when using EdgePier and compare it to a centralized container registry as the baseline.

\subsection{Experiment Setup}
For the following experiments, we host virtual machines on the Google Cloud platform with the following specifications: 2vCPU, 4GB memory and 20GB storage.

As the baseline, we use the official docker registry implementation \footnote{\url{https://docs.docker.com/registry/}}.
The centralized baseline registry and an EdgePier instance were deployed on a dedicated virtual machine that acted as the initial storage of images for the experiments. Furthermore, six virtual machines were hosted to emulate an edge site. 

The K3s\footnote{\url{https://k3s.io/}} Kubernetes distribution was installed as container orchestration platform and configured with the EdgePier prototype that is deployed on each node. Therefore, Pods started on each node were able to pull the respective image directly from the respective node.

In order to emulate a remote placement of the edge site, i.e. anywhere in a smart city, we limited the bandwidth on the registry node with the tc-NetEm\footnote{https://www.linux.org/docs/man8/tc-netem.html} tool to 500Mbps, 100Mbps, 50Mbps, and 20Mbps whereas the bandwidth between nodes in an edge site was not affected.

\subsection{Sequential starting of containers}
In this scenario, Pods based on the same 500MB docker image were scheduled one after another on each node of the infrastructure. This behavior represents the i.e. up-scaling of FaaS containers across a set of nodes or a rolling upgrade of an edge application. For this scenario we assumed a cold start of containers. Therefore, the image is not yet available on any node of the site. We ran the experiment ten times for each network bandwidth limitation and define the image distribution time as the time it took until each node of the site downloaded the full image. 

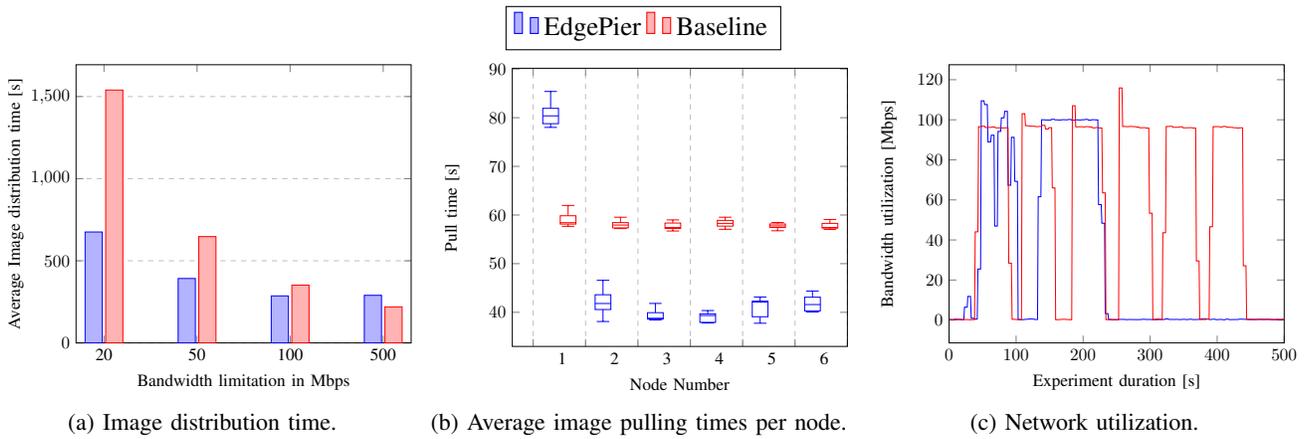
\begin{figure*}[ht]
\centering \ref{named}
\vspace*{1mm}
 
\subcaptionbox{Image distribution time.}
    {
        \begin{tikzpicture}[scale=0.65]
            \begin{axis}[
                ybar,
                xtick align=inside,
                xlabel={Bandwidth limitation in Mbps},
                ylabel={Average Image distribution time [s]},
                ymin=0, 
                symbolic x coords={20, 50, 100, 500},
                xtick=data,
                legend pos=north east,
                ymajorgrids=true,
                grid style=dashed,
                legend columns=-1,
                legend to name=named,
            ]
            \addplot table[x=limitation, y expr=\thisrow{ipfs} * 0.001, col sep=comma] {results/distribution_time_sequentially.csv}; \addlegendentry{EdgePier}
            \addplot table[x=limitation, y expr=\thisrow{normal} * 0.001, col sep=comma] {results/distribution_time_sequentially.csv}; \addlegendentry{Baseline}
            \end{axis}
        \end{tikzpicture}
    }
    \subcaptionbox{Average image pulling times per node.}
    {
         \begin{tikzpicture}[scale=0.65]
            \begin{axis}[
                boxplot/draw direction = y,
                ylabel={Pull time [s]},
                xlabel={Node Number},
                xmajorgrids,
                grid style=dashed,
                cycle list={{blue},{red}},
                boxplot={
                draw position={1/3 + floor(\plotnumofactualtype/2) + 1/3*mod(\plotnumofactualtype,2)},
                %
                box extend=0.3
                },
                xtick={0,1,2,...,6},
                xmax=6,
                x tick label as interval,
                xticklabels={
                    { 1},
                    { 2},
                    { 3},
                    { 4},
                    { 5},
                    { 6},
                },
            ]
            \addplot+ [boxplot]
            table [col sep=comma, y expr=\thisrow{ipfs-k3s-master} * 0.001] {results/boxplot-seq-100mbps.csv};
            \addplot+ [boxplot]
            table [col sep=comma, y expr=\thisrow{normal-k3s-master} * 0.001] {results/boxplot-seq-100mbps.csv};
            \addplot+ [boxplot]
            table [col sep=comma, y expr=\thisrow{ipfs-k3s-worker-1} * 0.001] {results/boxplot-seq-100mbps.csv};
            \addplot+ [boxplot]
            table [col sep=comma, y expr=\thisrow{normal-k3s-worker-1} * 0.001] {results/boxplot-seq-100mbps.csv};
            \addplot+ [boxplot]
            table [col sep=comma, y expr=\thisrow{ipfs-k3s-worker-2} * 0.001] {results/boxplot-seq-100mbps.csv};
            \addplot+ [boxplot]
            table [col sep=comma, y expr=\thisrow{normal-k3s-worker-2} * 0.001] {results/boxplot-seq-100mbps.csv};
            \addplot+ [boxplot]
            table [col sep=comma, y expr=\thisrow{ipfs-k3s-worker-3} * 0.001] {results/boxplot-seq-100mbps.csv};
            \addplot+ [boxplot]
            table [col sep=comma, y expr=\thisrow{normal-k3s-worker-3} * 0.001] {results/boxplot-seq-100mbps.csv};
            \addplot+ [boxplot]
            table [col sep=comma, y expr=\thisrow{ipfs-k3s-worker-4} * 0.001] {results/boxplot-seq-100mbps.csv};
            \addplot+ [boxplot]
            table [col sep=comma, y expr=\thisrow{normal-k3s-worker-4} * 0.001] {results/boxplot-seq-100mbps.csv};
            \addplot+ [boxplot]
            table [col sep=comma, y expr=\thisrow{ipfs-k3s-worker-5} * 0.001] {results/boxplot-seq-100mbps.csv};
            \addplot+ [boxplot]
            table [col sep=comma, y expr=\thisrow{normal-k3s-worker-5} * 0.001] {results/boxplot-seq-100mbps.csv};
            \end{axis}
        \end{tikzpicture}
    }
\subcaptionbox{Network utilization.}
    {
        \begin{tikzpicture}[scale=0.65]
            \begin{axis}[
                        xmin=0,
                        xmax=500,
                        xlabel = {Experiment duration [s]},
                        ylabel = {Bandwidth utilization [Mbps]},
                        no markers,
                        legend style={draw=none,
                            legend columns=-1,
                            anchor=south west,
                            legend cell align=left,
                             at={(0,1.00)},}
                        ]
                \addplot table [x=Time, y expr=\thisrow{EdgePier} / 8 / 125000, col sep=comma] {results/network_util_100mbps_sequen.csv};
                \addplot table [x=Time, y expr=\thisrow{Normal} / 8 / 125000, col sep=comma ] {results/network_util_100mbps_sequen.csv};
            \end{axis}
        \end{tikzpicture}
    }
    \caption{Sequentially starting a container on each node of the site, starting with Node 1. The subplots depict the average image distribution time for different bandwidth limitations to the cloud (a), the average pulling times per node (b) and the network utilization (c) on the registry node (both with 100mbps bandwidth limitation).}
    \label{fig:sequential}
\end{figure*}

As illustrated in Figure \ref{fig:sequential} (a), in a data center environment with 500 Mbps bandwidth between the nodes and the registry the official docker registry still performs significantly quicker than EdgePier. This is due to the overhead introduced by the IPFS DHT lookups necessary to locate image layers with EdgerPier. But as soon as the bandwidth to the registry instance is limited, i.e. to 100Mbps or 50Mbps, EdgePier is able to accelerate the image distribution, since the nodes starts sharing image data between themselves. In case of a 20Mbps bandwidth limitation, the image distribution time was even decreased by 56\% in our experiment.

Figure \ref{fig:sequential} (b) shows the average image download duration (pull time) per node with a 100Mbps bandwidth limitation to the registry: The image was first downloaded on Node 1 and then subsequently on the next nodes, in ascending order. As expected, the download duration remains constant for the baseline registry, since all container agents on the nodes are using the same uplink connection one after another. In contrast, the initial pulling on Node 1 still takes longer with EdgePier, however the following nodes are able to start the container faster due to the parallel synchronization of image data across the site.

Figure \ref{fig:sequential} (c) depicts the network utilization on the registry node, again with a 100Mbps limitation. As can be seen, the network layer is utilized to 100\% each time a pod is started on a node with the baseline registry, whereas for EdgePier the network layer is just utilized highly at the beginning of the experiment (for the initial download of the image into the edge site) and further download traffic is distributed across the other nodes in the site.

Summarizing, the results show that EdgePier is able to accelerate the image distribution process for remotely located edge sites. Use case scenarios in which containers are started one after another -- i.e. the aforementioned up-scaling of FaaS containers or rolling upgrades of an edge application -- greatly benefit from EdgePier, since as soon as one node of the edge sites downloaded the image it starts sharing the image with neighboring nodes. The network utilization is also decreased with EdgePier, therefore other and possible critical network transmissions to the cloud can make use of a higher bandwidth and subsequently lower response latencies.

\begin{figure*}[ht]
\centering \ref{named}
\vspace*{2mm}
 
\subcaptionbox{Image distribution time.}
    {
        \begin{tikzpicture}[scale=0.65]
            \begin{axis}[
                ybar,
                xtick align=inside,
                xlabel={Bandwidth limitation in Mbps},
                ylabel={Average Image distribution time [s]},
                ymin=0, 
                symbolic x coords={20, 50, 100, 500},
                xtick=data,
                ymajorgrids=true,
                grid style=dashed,
                legend columns=-1,
                legend to name=named,
            ]
            \addplot table[x=limitation, y expr=\thisrow{ipfs} * 0.001, col sep=comma] {results/distribution-time-concurrent.csv}; \addlegendentry{EdgePier}
            \addplot table[x=limitation, y expr=\thisrow{normal} * 0.001, col sep=comma] {results/distribution-time-concurrent.csv}; \addlegendentry{Baseline}
            \end{axis}
        \end{tikzpicture}
    }
\subcaptionbox{Average image pulling time per node.}
    {
         \begin{tikzpicture}[scale=0.65]
            \begin{axis}[
                boxplot/draw direction = y,
                ylabel={Pull time [s]},
                xlabel={Node Number},
                xmajorgrids,
                grid style=dashed,
                cycle list={{blue},{red}},
                boxplot={
                draw position={1/3 + floor(\plotnumofactualtype/2) + 1/3*mod(\plotnumofactualtype,2)},
                %
                box extend=0.3
                },
                xtick={0,1,2,...,6},
                xmax=6,
                x tick label as interval,
                xticklabels={
                    { 1},
                    { 2},
                    { 3},
                    { 4},
                    { 5},
                    { 6},
                },
            ]
            \addplot+ [boxplot]
            table [col sep=comma, y expr=\thisrow{ipfs-k3s-master} * 0.001] {results/boxplot-concurrent-100mbps.csv};
            \addplot+ [boxplot]
            table [col sep=comma, y expr=\thisrow{normal-k3s-master} * 0.001] {results/boxplot-concurrent-100mbps.csv};
            \addplot+ [boxplot]
            table [col sep=comma, y expr=\thisrow{ipfs-k3s-worker-1} * 0.001] {results/boxplot-concurrent-100mbps.csv};
            \addplot+ [boxplot]
            table [col sep=comma, y expr=\thisrow{normal-k3s-worker-1} * 0.001] {results/boxplot-concurrent-100mbps.csv};
            \addplot+ [boxplot]
            table [col sep=comma, y expr=\thisrow{ipfs-k3s-worker-2} * 0.001] {results/boxplot-concurrent-100mbps.csv};
            \addplot+ [boxplot]
            table [col sep=comma, y expr=\thisrow{normal-k3s-worker-2} * 0.001] {results/boxplot-concurrent-100mbps.csv};
            \addplot+ [boxplot]
            table [col sep=comma, y expr=\thisrow{ipfs-k3s-worker-3} * 0.001] {results/boxplot-concurrent-100mbps.csv};
            \addplot+ [boxplot]
            table [col sep=comma, y expr=\thisrow{normal-k3s-worker-3} * 0.001] {results/boxplot-concurrent-100mbps.csv};
            \addplot+ [boxplot]
            table [col sep=comma, y expr=\thisrow{ipfs-k3s-worker-4} * 0.001] {results/boxplot-concurrent-100mbps.csv};
            \addplot+ [boxplot]
            table [col sep=comma, y expr=\thisrow{normal-k3s-worker-4} * 0.001] {results/boxplot-concurrent-100mbps.csv};
            \addplot+ [boxplot]
            table [col sep=comma, y expr=\thisrow{ipfs-k3s-worker-5} * 0.001] {results/boxplot-concurrent-100mbps.csv};
            \addplot+ [boxplot]
            table [col sep=comma, y expr=\thisrow{normal-k3s-worker-5} * 0.001] {results/boxplot-concurrent-100mbps.csv};
            \end{axis}
        \end{tikzpicture}
    }
\subcaptionbox{Network utilization}
    {
        \begin{tikzpicture}[scale=0.65]
            \begin{axis}[
                        xmin=0,
                        xmax=400,
                        xlabel = {Experiment duration [s]},
                        ylabel = {Bandwidth utilization [Mbps]},
                        no markers,
                        legend style={draw=none,
                            legend columns=-1,
                            anchor=south west,
                            legend cell align=left,
                             at={(0,1.00)},}
                        ]
                \addplot table [x=Time, y expr=\thisrow{EdgePier} / 8 / 125000, col sep=comma] {results/network_util_100mbps_concurrent.csv};
                \addplot table [x=Time, y expr=\thisrow{Normal} / 8 / 125000 , col sep=comma ] {results/network_util_100mbps_concurrent.csv};
            \end{axis}
        \end{tikzpicture}
    }
    \caption{Concurrently starting containers on each node of the infrastructure. The image distribution time (a) refers to the longest pulling time of the image in the site and the average pulling times per node (b) and the network utilization (c) are shown for a 100Mbps bandwidth limitation to the registry node. }
    \label{fig:concurrent}
\end{figure*}
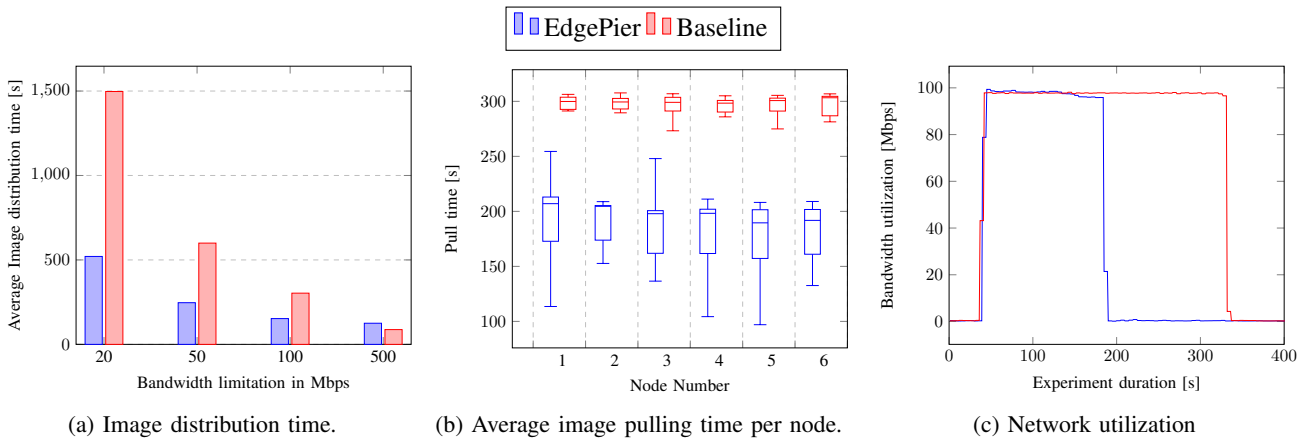

\subsection{Concurrent starting of containers}
For this experiment, a Pod was started on each node of the edge site at the same time, resulting into concurrent pulling of images to the site. This is often the case when i.e. ML models need to be distributed to each node of the site, or an edge application is updated by for instance a continuous delivery pipeline. 
Here, the full distribution time is defined as time until each node has successfully pulled the image and started the container, thus it refers to the longest pulling time of the image in the site.

We were using the same 500MB image as in the sequential scenario and also repeated the experiment ten times for each bandwidth limitation.

Figure \ref{fig:concurrent} (a) depicts the image distribution time and shows similar behaviour to the sequential scenario: With a high bandwidth images can distributed a bit faster to all nodes when using the baseline registry, since in a data center environment the network link can be fully utilized by all container agents. However, with limited bandwidth
EdgePier is able to accelerate the image distribution process by up to 65\% (20Mbps bandwidth limitation).

As can be seen in Figure \ref{fig:concurrent} (b), the standard deviation of pull times (also referred as download duration) per node is significantly higher for EdgePier compared to the baseline registry or previous scenario, since in this case all nodes start pulling the image at the same time. This is due to the fact that layers of a docker image are not necessarily pulled in successive order and nodes of the site start sharing the layer blobs with other nodes as soon as they successfully downloaded them. Subsequently, it resulted in variations of the download duration during the experiment because i.e. sometimes a layer was downloaded by several nodes at the same time.

Nevertheless, the average download duration per node are still faster for EdgePier when limiting the bandwidth to 100Mbps as shown in Figure \ref{fig:concurrent} (b), since although the image pulling process was started at the same time, nodes were able to also download layers from other nodes after some time.

This can also be seen in Figure \ref{fig:concurrent} (c) which depicts the network utilization on the uplink connection to the registry service when limiting the bandwidth to 100Mbps: Although for both cases the network link is fully utilized for the whole experiment duration, EdgePier only requires just half of the time to distribute images to every node of the site, and therefore also only occupies the network link for half the time. Consequently, the network link is exonerated when using EdgePier, which leaves room for other important Edge-Cloud communication on the uplink.

Summarizing, the results show that even with high concurrent deployment load in the edge site, EdgePier was able to significantly improve the image provisioning time while also decreasing the network load. Therefore, the EdgePier approach can greatly benefit use cases in which containers need to be started across an edge site regularly, i.e. when an updated application or model is deployed. 

Furthermore, in this experiment we evaluated the worst case scenario since no layer of the image was already available in the edge site. If some of the image layers are already available in the site, for instance because the same base image was used by another container, we expect that the performance is even further enhanced, due to the deduplication offered by Docker and IPFS.

\subsection{Impact of image size}
As a third experiment we evaluated the impact of the image size on the average pull time per node. The two previous experiments were conducted with a 500MB container image, and in order to analyze how EdgePier performs with different images sizes we created ten docker images from 100MB to 1GB. Therefore, we added a new layer with a randomized 100MB binary file for each new image.

Afterwards, we concurrently started the image pulling process on each node of the site and monitored the average pulling time per node for EdgePier and the centralized registry as a baseline. The experiment was repeated ten times for each image size and the bandwidth to the cloud was limited to 100Mbps to simulate a remote location of the site. After each iteration, the image was deleted from the nodes and local IPFS storage, resulting in a cold start of the image for each experiment run.

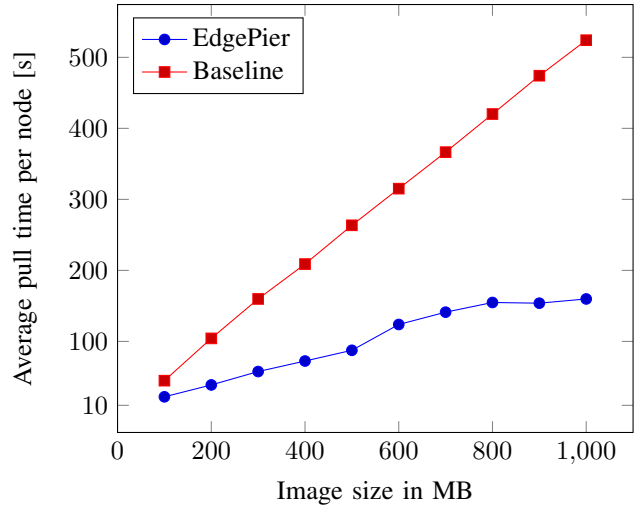
\begin{figure}
    \centering
            \begin{tikzpicture}
            \begin{axis}[
                        xmin=0,
                        xmax=1100,
                        ytick = {10,100,200,...,500},
                        xlabel = {Image size in MB},
                        ylabel = {Average pull time per node [s]},
                    legend cell align={left},
                    legend pos=north west,
                        ]
                \addplot table [x=ImageSize, y expr=\thisrow{EdgePier} * 0.001, col sep=comma] {results/different-sizes.csv};
                \addlegendentry{EdgePier}
                \addplot table [x=ImageSize, y expr=\thisrow{Normal} * 0.001, col sep=comma ] {results/different-sizes.csv};
                \addlegendentry{Baseline}
            \end{axis}
        \end{tikzpicture}
    \caption{Average pull time per node for different container image sizes. The images were concurrently downloaded on six nodes of an edge site and the bandwdith was limited to 100Mbps.}
    \label{fig:sizes}
\end{figure}

As can be seen in Figure \ref{fig:sizes}, the average pull time per node increases linearly for the baseline registry with increasing images sizes, since all nodes are using the the same uplink connection to the cloud during the download. As a result, the download time increases for larger images. In case of EdgePier, the average download duration per node increases slower, because nodes start sharing image layers and therefore relieving the network uplink. For instance, between 800MB and 1GB the average download time is only marginally increased by up to 5 seconds. This is due to the fact that with growing image sizes the amount of layers, that can possibly be shared by nodes, also increases. As mentioned before, image layers are not downloaded in order by the container runtime agent and accordingly it is more likely that nodes download different layers during the concurrent pulling, which then do not need to be downloaded again via the uplink connection.

Summarizing, the results confirm the previous observed improvements offered by EdgePier for the image provisioning to remote edge sites: For small images of i.e. 100MB and 200MB size, EdgePier already performs better than the baseline registry, although the decrease in the average pull time per node is still relatively small. However, with increasing image sizes the decentralized peer-to-peer registry EdgePier provides significantly improved image provisioning times compared to a centralized registry.

Consequently, EdgePier is able to offer enhancements for a wide range of container deployments to remote edge sites, regardless of the image sizes. Therefore, it fulfills the previous stated requirement for an \textit{accelerated image distribution}.

%

\section{Related Work}
In this section we present and review related work in the area of container image distribution. Although some of the related approaches also utilize peer-to-peer connections to distribute images between nodes, they mostly focus on BitTorrent-based technologies:
Kangjin et al \cite{kangjinFIDFasterImage2017} present a P2P registry called FID, that distributes the layers of an image via torrents. They provide an architecture based on several torrent trackers and a FID agent that is deployed on every node of the infrastructure. The envisioned environments for their approach are large-scale data centers and in contrast to EdgePier, FID is not fully decentralized since it relies on the synchronized tracker services. In addition, the tracker details are stored in each torrent file, thus they need to be recreated when a trackers fail or are scaled up.

In the industry, similar approaches were taken by Uber respectively Alibaba: Uber developed the P2P registry Kraken\footnote{https://github.com/uber/kraken}, which utilizes dedicated seeder nodes, trackers and metadata stores based on a distributed file system for the container image management and deploys agents on each node of the infrastructure. 
The Dragonfly\footnote{https://github.com/dragonflyoss/Dragonfly} registry service  
provided by Alibaba also employs several dedicated nodes for the coordination of image data transfers. Both registries leverage protocols based on BitTorrent and focus on cloud data centers where they show promising results, nevertheless they depend on resource intensive dedicated services that introduce single point of failures and might overload small edge sites.

More similar to our approach is the work by Gazzetti et al. \cite{gazzettiScalableLinuxContainer2018}
that presents a stream based peer-to-peer image provisioning for edge computing platforms. The authors describe an architecture consisting of a gateway manager which download images on behalf of an edge site and coordinates the distribution to stream managers which are deployed on every node. 

Another area of the related work aims to improve the image provisioning process by employing distributed file systems: Nathan et al. \cite{nathanCoMIConCoOperativeManagement2017} adjusted the docker daemon and implemented cooperative downloading of partial images to a shared storage. The daemons simultaneously download a requested image and are coordinated by a provisioning and image distribution manager. A similar method was proposed by Du et al. \cite{duCiderRapidDocker2017}, where Ceph is leveraged as the underlying storage for docker images. In addition, the authors introduced a copy-on-write layer for containers in order to ensure scalability during high deployment load.

The Wharf middleware \cite{zhengWharfSharingDocker2018} is also able to load image layers in parallel with a set of docker daemons to tackle the problem of redundant pulls. Therefore, Zheng et al. adjusted the docker code base and present a synchronization mechanism to store images on a distributed file system. The store layer is interchangeable, and Wharf can work with any POSIX compatible storage layer.

In contrast to the EdgePier approach, the aforementioned methods are mainly focused on cloud data centers, where distributed file systems usually offer a good performance. However, in Edge and Fog computing environments the unreliable network, lightweight and dynamic nature of devices as well as node churns often complicate the usage of distributed storage. Therefore, we employ a decentralized peer-to-peer approach which is not dependent on centralized or dedicated services and can cope with constantly changing environments. 

\section{Conclusion}
In this paper we presented the peer-to-peer container image registry EdgePier, which can be deployed in a fully decentralized manner. Nodes inside remotely located edge site share image layers with each other, resulting in accelerated container provisioning times in case of limited edge-cloud uplink connections.

The results of the conducted evaluation show that EdgePier is able to outperform a centralized baseline registry by up to 65\% in terms of image distribution time. At the same time the network connection to the cloud is relieved, which improves other and potentially critical network transmissions between the edge sites and the cloud.

Summarizing, EdgePier shows promising results for highly distributed edge sites, enables fast application deployments and therefore benefits a variety of use cases in edge and fog computing, smart cities and IoT in general.

In future work we plan to conduct further research on cross-site image exchange to i.e. enable a model exchange between edge sites.

\bibliographystyle{IEEEtran}
\bibliography{p2p}

\begin{thebibliography}{10}
\providecommand{\url}[1]{#1}
\csname url@samestyle\endcsname
\providecommand{\newblock}{\relax}
\providecommand{\bibinfo}[2]{#2}
\providecommand{\BIBentrySTDinterwordspacing}{\spaceskip=0pt\relax}
\providecommand{\BIBentryALTinterwordstretchfactor}{4}
\providecommand{\BIBentryALTinterwordspacing}{\spaceskip=\fontdimen2\font plus
\BIBentryALTinterwordstretchfactor\fontdimen3\font minus
  \fontdimen4\font\relax}
\providecommand{\BIBforeignlanguage}[2]{{%
\expandafter\ifx\csname l@#1\endcsname\relax
\typeout{** WARNING: IEEEtran.bst: No hyphenation pattern has been}%
\typeout{** loaded for the language `#1'. Using the pattern for}%
\typeout{** the default language instead.}%
\else
\language=\csname l@#1\endcsname
\fi
#2}}
\providecommand{\BIBdecl}{\relax}
\BIBdecl

\bibitem{ismailEvaluationDockerEdge2015b}
B.~I. Ismail, E.~Mostajeran~Goortani, M.~B. Ab~Karim, W.~Ming~Tat, S.~Setapa,
  J.~Y. Luke, and O.~Hong~Hoe, ``\BIBforeignlanguage{en}{Evaluation of
  {{Docker}} as {{Edge}} computing platform},'' in
  \emph{\BIBforeignlanguage{en}{2015 {{IEEE Conference}} on {{Open Systems}}
  ({{ICOS}})}}.\hskip 1em plus 0.5em minus 0.4em\relax {Bandar Melaka}: {IEEE},
  Aug. 2015, pp. 130--135.

\bibitem{sharmaContainersVirtualMachines2016}
P.~Sharma, L.~Chaufournier, P.~Shenoy, and Y.~C. Tay,
  ``\BIBforeignlanguage{en}{Containers and {{Virtual Machines}} at {{Scale}}:
  {{A Comparative Study}}},'' in \emph{\BIBforeignlanguage{en}{Proceedings of
  the 17th {{International Middleware Conference}}}}.\hskip 1em plus 0.5em
  minus 0.4em\relax {Trento Italy}: {ACM}, Nov. 2016, pp. 1--13.

\bibitem{morabitoHypervisorsVsLightweight2015}
R.~Morabito, J.~Kjallman, and M.~Komu, ``\BIBforeignlanguage{en}{Hypervisors
  vs. {{Lightweight Virtualization}}: {{A Performance Comparison}}},'' in
  \emph{\BIBforeignlanguage{en}{2015 {{IEEE International Conference}} on
  {{Cloud Engineering}}}}.\hskip 1em plus 0.5em minus 0.4em\relax {Tempe, AZ,
  USA}: {IEEE}, Mar. 2015, pp. 386--393.

\bibitem{bellavistaFeasibilityFogComputing2017b}
P.~Bellavista and A.~Zanni, ``\BIBforeignlanguage{en}{Feasibility of {{Fog
  Computing Deployment}} based on {{Docker Containerization}} over
  {{RaspberryPi}}},'' in \emph{\BIBforeignlanguage{en}{Proceedings of the 18th
  {{International Conference}} on {{Distributed Computing}} and
  {{Networking}}}}.\hskip 1em plus 0.5em minus 0.4em\relax {Hyderabad India}:
  {ACM}, Jan. 2017, pp. 1--10.

\bibitem{al-rakhamiLightweightCostEffective2020}
M.~{Al-Rakhami}, A.~Gumaei, M.~Alsahli, M.~M. Hassan, A.~Alamri, A.~Guerrieri,
  and G.~Fortino, ``\BIBforeignlanguage{en}{A lightweight and cost effective
  edge intelligence architecture based on containerization technology},''
  \emph{\BIBforeignlanguage{en}{World Wide Web}}, vol.~23, no.~2, pp.
  1341--1360, Mar. 2020.

\bibitem{hallExecutionModelServerless2019}
A.~Hall and U.~Ramachandran, ``\BIBforeignlanguage{en}{An execution model for
  serverless functions at the edge},'' in
  \emph{\BIBforeignlanguage{en}{Proceedings of the {{International Conference}}
  on {{Internet}} of {{Things Design}} and {{Implementation}}}}.\hskip 1em plus
  0.5em minus 0.4em\relax {Montreal Quebec Canada}: {ACM}, Apr. 2019, pp.
  225--236.

\bibitem{gliksonDevicelessEdgeComputing2017}
A.~Glikson, S.~Nastic, and S.~Dustdar, ``\BIBforeignlanguage{en}{Deviceless
  edge computing: Extending serverless computing to the edge of the network},''
  in \emph{\BIBforeignlanguage{en}{Proceedings of the 10th {{ACM International
  Systems}} and {{Storage Conference}}}}.\hskip 1em plus 0.5em minus
  0.4em\relax {Haifa Israel}: {ACM}, May 2017, pp. 1--1.

\bibitem{baldiniServerlessComputingCurrent2017a}
I.~Baldini, P.~Castro, K.~Chang, P.~Cheng, S.~Fink, V.~Ishakian, N.~Mitchell,
  V.~Muthusamy, R.~Rabbah, A.~Slominski, and P.~Suter,
  ``\BIBforeignlanguage{en}{Serverless {{Computing}}: {{Current Trends}} and
  {{Open Problems}}},'' \emph{\BIBforeignlanguage{en}{arXiv:1706.03178 [cs]}},
  Jun. 2017.

\bibitem{mannerColdStartInfluencing2018}
J.~Manner, M.~Endre{\ss}, T.~Heckel, and G.~Wirtz, ``Cold {{Start Influencing
  Factors}} in {{Function}} as a {{Service}},'' in \emph{2018 {{IEEE}}/{{ACM
  International Conference}} on {{Utility}} and {{Cloud Computing Companion}}
  ({{UCC Companion}})}, Dec. 2018, pp. 181--188.

\bibitem{kangjinFIDFasterImage2017}
W.~Kangjin, Y.~Yong, L.~Ying, L.~Hanmei, and M.~Lin,
  ``\BIBforeignlanguage{en}{{{FID}}: {{A Faster Image Distribution System}} for
  {{Docker Platform}}},'' in \emph{\BIBforeignlanguage{en}{2017 {{IEEE}} 2nd
  {{International Workshops}} on {{Foundations}} and {{Applications}} of
  {{Self}}* {{Systems}} ({{FAS}}*{{W}})}}.\hskip 1em plus 0.5em minus
  0.4em\relax {Tucson, AZ, USA}: {IEEE}, Sep. 2017, pp. 191--198.

\bibitem{zhengWharfSharingDocker2018}
C.~Zheng, L.~Rupprecht, V.~Tarasov, D.~Thain, M.~Mohamed, D.~Skourtis, A.~S.
  Warke, and D.~Hildebrand, ``\BIBforeignlanguage{en}{Wharf: {{Sharing Docker
  Images}} in a {{Distributed File System}}},'' in
  \emph{\BIBforeignlanguage{en}{Proceedings of the {{ACM Symposium}} on {{Cloud
  Computing}}}}.\hskip 1em plus 0.5em minus 0.4em\relax {Carlsbad CA USA}:
  {ACM}, Oct. 2018, pp. 174--185.

\bibitem{nathanCoMIConCoOperativeManagement2017}
S.~Nathan, R.~Ghosh, T.~Mukherjee, and K.~Narayanan,
  ``\BIBforeignlanguage{en}{{{CoMICon}}: {{A Co}}-{{Operative Management
  System}} for {{Docker Container Images}}},'' in
  \emph{\BIBforeignlanguage{en}{2017 {{IEEE International Conference}} on
  {{Cloud Engineering}} ({{IC2E}})}}.\hskip 1em plus 0.5em minus 0.4em\relax
  {Vancouver, BC, Canada}: {IEEE}, Apr. 2017, pp. 116--126.

\bibitem{duCiderRapidDocker2017}
L.~Du, T.~Wo, R.~Yang, and C.~Hu, ``\BIBforeignlanguage{en}{Cider: A {{Rapid
  Docker Container Deployment System}} through {{Sharing Network Storage}}},''
  in \emph{\BIBforeignlanguage{en}{2017 {{IEEE}} 19th {{International
  Conference}} on {{High Performance Computing}} and {{Communications}};
  {{IEEE}} 15th {{International Conference}} on {{Smart City}}; {{IEEE}} 3rd
  {{International Conference}} on {{Data Science}} and {{Systems}}
  ({{HPCC}}/{{SmartCity}}/{{DSS}})}}.\hskip 1em plus 0.5em minus 0.4em\relax
  {Bangkok}: {IEEE}, Dec. 2017, pp. 332--339.

\bibitem{littleyBoltScalableDocker2019a}
M.~Littley, A.~Anwar, H.~Fayyaz, Z.~Fayyaz, V.~Tarasov, L.~Rupprecht,
  D.~Skourtis, M.~Mohamed, H.~Ludwig, Y.~Cheng, and A.~R. Butt,
  ``\BIBforeignlanguage{en}{Bolt: {{Towards}} a {{Scalable Docker Registry}}
  via {{Hyperconvergence}}},'' in \emph{\BIBforeignlanguage{en}{2019 {{IEEE}}
  12th {{International Conference}} on {{Cloud Computing}} ({{CLOUD}})}}.\hskip
  1em plus 0.5em minus 0.4em\relax {Milan, Italy}: {IEEE}, Jul. 2019, pp.
  358--366.

\bibitem{benetIPFSContentAddressed2014}
J.~Benet, ``{{IPFS}} - {{Content Addressed}}, {{Versioned}}, {{P2P File
  System}},'' \emph{arXiv:1407.3561 [cs]}, Jul. 2014.

\bibitem{maymounkovKademliaPeertoPeerInformation2002}
P.~Maymounkov and D.~Mazi{\`e}res, ``\BIBforeignlanguage{en}{Kademlia: {{A
  Peer}}-to-{{Peer Information System Based}} on the {{XOR Metric}}},'' in
  \emph{\BIBforeignlanguage{en}{Peer-to-{{Peer Systems}}}}, ser. Lecture
  {{Notes}} in {{Computer Science}}, P.~Druschel, F.~Kaashoek, and A.~Rowstron,
  Eds.\hskip 1em plus 0.5em minus 0.4em\relax {Berlin, Heidelberg}: {Springer},
  2002, pp. 53--65.

\bibitem{10.1007/3-540-48184-2_32}
R.~C. Merkle, ``A digital signature based on a conventional encryption
  function,'' in \emph{Advances in Cryptology \textemdash{} {{CRYPTO}} '87},
  C.~Pomerance, Ed.\hskip 1em plus 0.5em minus 0.4em\relax {Berlin,
  Heidelberg}: {Springer Berlin Heidelberg}, 1988, pp. 369--378.

\bibitem{hoqueContainerOrchestrationFog2017a}
S.~Hoque, M.~S. De~Brito, A.~Willner, O.~Keil, and T.~Magedanz,
  ``\BIBforeignlanguage{en}{Towards {{Container Orchestration}} in {{Fog
  Computing Infrastructures}}},'' in \emph{\BIBforeignlanguage{en}{2017
  {{IEEE}} 41st {{Annual Computer Software}} and {{Applications Conference}}
  ({{COMPSAC}})}}.\hskip 1em plus 0.5em minus 0.4em\relax {Turin, Italy}:
  {IEEE}, Jul. 2017, pp. 294--299.

\bibitem{gazzettiScalableLinuxContainer2018}
M.~Gazzetti, A.~Reale, K.~Katrinis, and A.~Corradi,
  ``\BIBforeignlanguage{en}{Scalable {{Linux Container Provisioning}} in
  {{Fog}} and {{Edge Computing Platforms}}},'' in
  \emph{\BIBforeignlanguage{en}{Euro-{{Par}} 2017: {{Parallel Processing
  Workshops}}}}, D.~B. Heras, L.~Boug{\'e}, G.~Mencagli, E.~Jeannot,
  R.~Sakellariou, R.~M. Badia, J.~G. Barbosa, L.~Ricci, S.~L. Scott, S.~Lankes,
  and J.~Weidendorfer, Eds.\hskip 1em plus 0.5em minus 0.4em\relax {Cham}:
  {Springer International Publishing}, 2018, vol. 10659, pp. 304--315.

\end{thebibliography}
\end{document}